\documentclass[aps,prl,showpacs,twocolumn,superscriptaddress,floatfix,tightenlines]{revtex4}
\pdfoutput=1 
\usepackage{bbm}
\usepackage{mathrsfs}
\usepackage{epsfig,psfrag}
\usepackage{amsmath,amsfonts,amssymb}
\usepackage[usenames]{color}
\usepackage{bm}

\newcommand{\cE}{\mathcal E}

\newcommand{\tr}{\text{Tr}}    

\definecolor{BrickRed}{cmyk}{0,0.89,0.94,0.28}
\definecolor{MidnightBlue}{cmyk}{0.98,0.13,0,0.43}
\definecolor{DarkGreen}{rgb}{0,0.7,0.1}

\begin{document}

\preprint{draft}

\title{Three-body Casimir effects and non-monotonic forces}

\author{P.\ Rodriguez-Lopez}
\affiliation{Departamento de Fisica Aplicada I, Universidad Complutense, 28040 Madrid, Spain}

\author{S.\ J.\ Rahi}
\affiliation{Massachusetts Institute of Technology, Department of
  Physics, 77 Massachusetts Avenue, Cambridge, MA 02139, USA}

\author{T.\ Emig}
\affiliation{Institut f\"ur Theoretische Physik, Universit\"at zu K\"oln,
Z\"ulpicher Strasse 77, 50937 K\"oln, Germany}
\affiliation{Laboratoire de Physique Th\'eorique et Mod\`eles
Statistiques, CNRS UMR 8626, Universit\'e Paris-Sud, 91405 Orsay,
France}

\date{\today}

\begin{abstract}
  Casimir interactions are not pair-wise additive. This property leads
  to collective effects that we study for a pair of objects near a
  conducting wall. We employ a scattering approach to compute the
  interaction in terms of fluctuating multipoles. The wall can lead to
  a non-monotonic force between the objects. For two atoms with
  anisotropic electric and magnetic dipole polarizabilities we
  demonstrate that this non-monotonic effect results from a
  competition between two- and three body interactions. By including
  higher order multipoles we obtain the force between two macroscopic
  metallic spheres for a wide range of sphere separations and
  distances to the wall.
\end{abstract}

\pacs{42.25.Fx, 03.70.+k, 12.20.-m.}

\maketitle


A hallmark property of dispersion forces is their non-additivity which
clearly distinguishes them from electromagnetic forces between charged
particles \cite{Parsegian:2005eu}.  Work on the interactions between
multiple objects is limited mostly to atoms or small particles which
are described well in dipole approximation \cite{Power:1982cs}.  This
approximation cannot be used for macroscopic objects at separations
that are comparable to their size since higher order multipole
fluctuations have to be included
\cite{Feinberg:1974xw,Emig:2007os}. In such situations, also other
common ``additive'' methods such as proximity or two-body-interaction
approximations fail.  Three-body effects for macroscopic bodies have
been studied in quasi two-dimensional (2D) geometries that are
composed of parallel perfect metal cylinders of quadratic
\cite{Rodriguez:2007a} or circular \cite{Rahi:2008bv,Rahi:2008kb}
cross section and parallel sidewalls.  For this setup non-monotonic
forces have been found and interpreted as resulting from a competition
between electric and magnetic polarizations which are decoupled for
quasi 2D geometries of perfect metal structures. In this Letter we
investigate collective 3-body effects between {\it compact} objects,
including anisotropic polarizabilities, and a wall in three dimensions
using a recently developed scattering approach
\cite{Emig:2007os,Emig:2008ee}. This allows us to observe the
influence of polarization coupling and anisotropy on non-monotonic
effects.

\begin{figure}[htbp]
\includegraphics[width=0.47\linewidth]{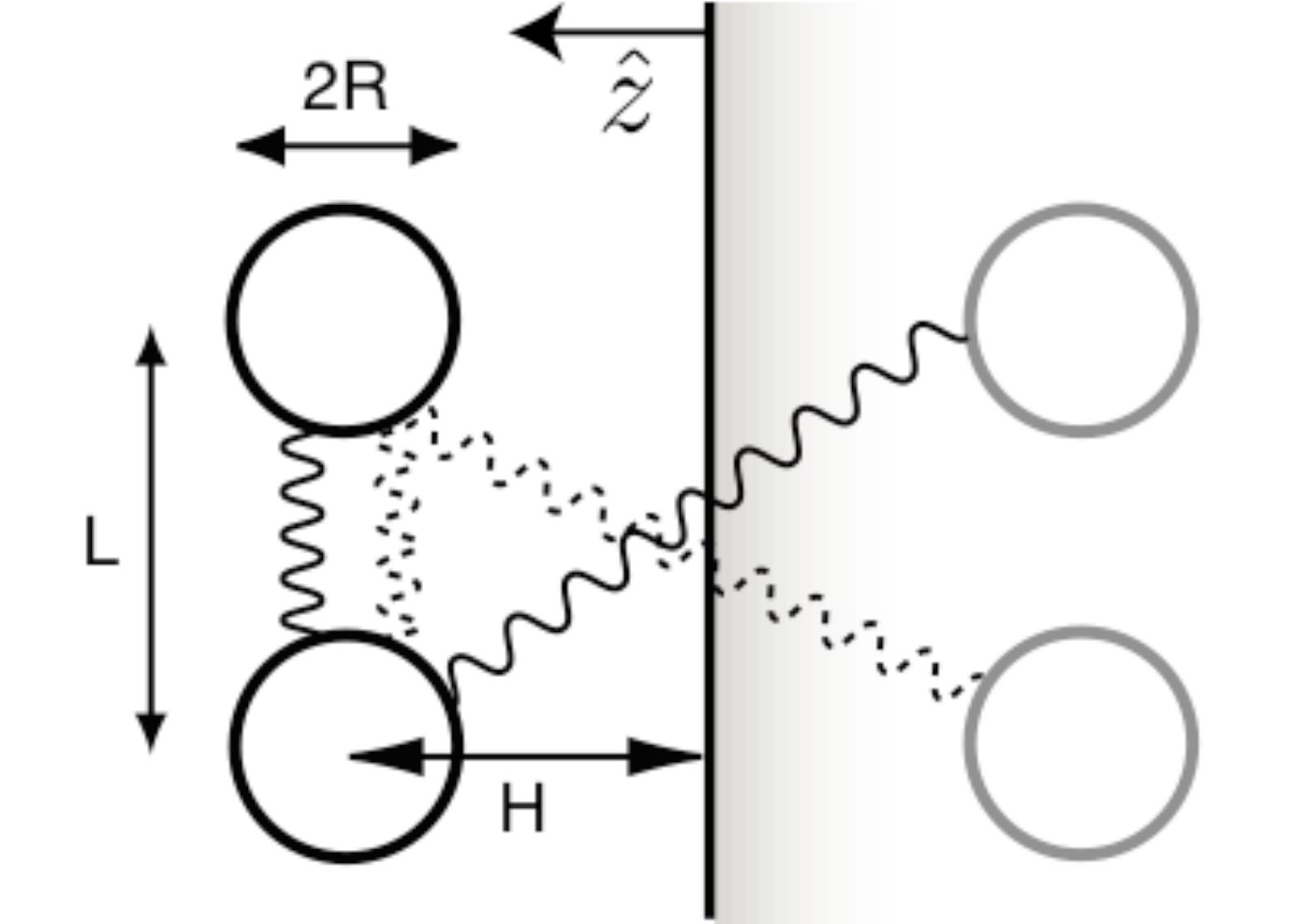}
  \caption{\label{fig:1}Geometry of the two-sphere/atom and sidewall
    system. Shown are also the mirror images (grey) and two- and
    three-body contributions (solid and dashed curly lines, respectively).}
\end{figure}

We consider the retarded Casimir interaction between a pair of atoms
with anisotropic electric and magnetic polarizabilities near a
conducting wall, see Fig.~\ref{fig:1}. We identify a competition
between 2- and 3-body effects and prove that this leads to a
non-monotonic dependence of the force between the atoms on the wall
separation $H$ for {\it each} of the four possible polarizations of
fluctuations (electric/magnetic and parallel/perpendicular to the
wall) {\it separately}. For isotropic polarizabilities we find that
only the force component due to electric fluctuations is non-monotonic
in $H$. This findings suggest the possibility to engineer the
monotonicity properties of the force by suitable tailoring of the
polarization tensors via the shape and material composition of
macroscopic particles.

For atoms, magnetic effects are almost always rather small in the
retarded limit.  Contrary to this, for conducting macroscopic objects
contributions from electric and magnetic multipole fluctuations are
comparable. To study the effect of higher-order
multipoles, we consider also two perfect metal spheres near a wall,
see Fig.~\ref{fig:1}.  Based on consistent analytical results for
large separations and numerical computations at smaller distances we
find a non-monotonic dependence of the force between the spheres on
$H$. Unlike for atoms, this effect is occurs at sufficiently
large sphere separations only.
\vspace*{-0.3cm}


As derived by Refs.~\cite{Emig:2007os,Emig:2008ee}, the Casimir energy
of two bodies in the presence of a perfectly conducting sidewall can
be obtained within a scattering approach by employing the method of
images that introduces fluctuating currents on the mirror bodies. The
Casimir energy of the original system is then given by the energy of
the original and the image objects and it can be expressed as
an integral over imaginary wave number,
\begin{equation}
  \label{eq:energy}
  \cE = \frac{\hbar c}{2\pi} \int_0^\infty d\kappa \,\ln \det ( {\mathbb M}\, {\mathbb M}_\infty^{-1}) 
\end{equation}
with the matrix
\begin{equation}
  \label{eq:M-matrix}
  {\mathbb M} = \begin{pmatrix}
 {\mathbb T}^{-1} + {\mathbb U}^{I,11} &  {\mathbb U}^{12} + {\mathbb U}^{I,12} \\
{\mathbb U}^{21} + {\mathbb U}^{I,21} &  {\mathbb T}^{-1} + {\mathbb U}^{I,22} 
\end{pmatrix} \, ,
\end{equation}
which is given by the T-matrix ${\mathbb T}$ that relates the
incomming and scattered electromagnetic (EM) fields for each body, and
by the U-matrices ${\mathbb U}^{\alpha\beta}$, ${\mathbb
  U}^{I,\alpha\beta}$ that describe the interaction between the
multipoles of object $\alpha$ and object $\beta$ and between the
multipoles of object $\alpha$ and the image of object $\beta$,
respectively. The T-matrix depends only on properties of the bodies
such as polarizability or size and shape. The U-matrices depend only
on the distance vector between the objects and decay exponentially
with distance and wave number $\kappa$.  The matrix ${\mathbb
  M}_\infty$ accounts for the subtraction of the object's
self-energies and hence follows from ${\mathbb M}$ by taking the limit
of infinite separations, i.e., by setting all U-matrices to zero. For
a multipole expansion the matrix elements are computed in a vector
spherical basis for the EM field with partial wave numbers $l\ge 1$,
$m=-l,\ldots, l$. (Details of this expansion and the U-matrix elements
can be found in Ref.~\cite{Emig:2008ee}.)
\begin{figure}[ht]
\includegraphics[width=1.\linewidth]{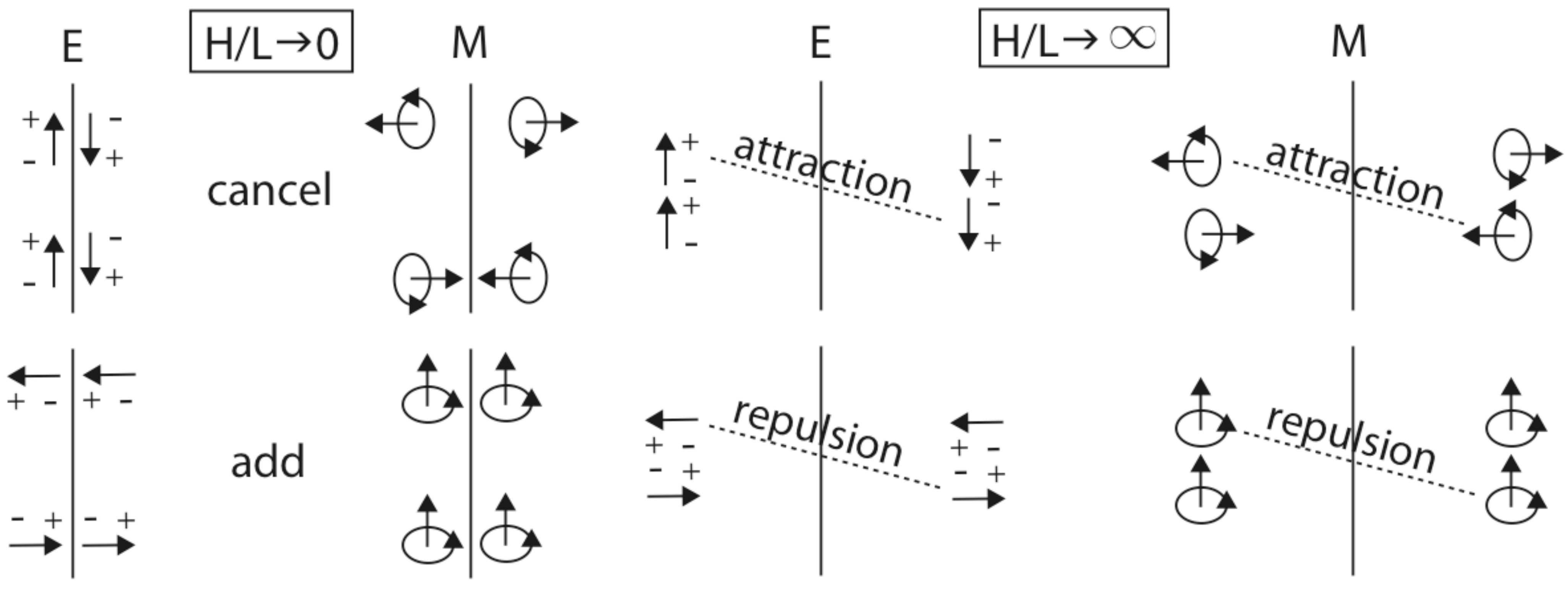}
\caption{\label{fig:fig2}Typical orientations of electric (E) and
  magnetic (M) dipoles and image dipoles for $H/L\to 0$ and
  $H/L\to\infty$.}
\end{figure}
\vspace*{-0.5cm}

In the following we study the force $F = -\partial \cE/\partial L$
between the two objects at separation $L$ and hence eliminate the
contributions to the energy that depend only on the sidewall
separation $H$, see Fig.~\ref{fig:1}. We expand the determinant of Eq.~\eqref{eq:energy} as
\begin{widetext}
\vspace*{-0.7cm}
\begin{equation}
  \label{eq:det-M-1}
\det ( {\mathbb M}\, {\mathbb M}_\infty^{-1}) = 
\det ( 1 +  {\mathbb T} {\mathbb U}^{I} )
\det ( 1 +  {\mathbb T} {\mathbb U}^{I} )
\det \left[ 1- ( 1 +  {\mathbb T} {\mathbb U}^{I} )^{-1}
  {\mathbb T} ( {\mathbb U}^{21} + {\mathbb U}^{I,21} ) 
 ( 1 +  {\mathbb T} {\mathbb U}^{I} )^{-1} 
  {\mathbb T} ( {\mathbb U}^{12} + {\mathbb U}^{I,12} ) \right] \, .
\end{equation}
The first two determinants on the r.h.s. yield together twice the
interaction energy between a single object and the sidewall since
${\mathbb U}^{I}\equiv{\mathbb U}^{I,11}= {\mathbb U}^{I,22}$
describes the multipole coupling between one object and its image and
hence depends only on $H$. Hence, we consider only the energy
$\cE_{\underline{\circ\circ}}$ that corresponds to the last
determinant of Eq.~\eqref{eq:det-M-1} and provides the potential
energy of the two objects in the presence of the sidewall so that
$F=-\partial \cE_{\underline{\circ\circ}}/\partial L$. In the absence
of the sidewall, $H\to\infty$, the matrices ${\mathbb
  U}^{I,\alpha\beta}$ all vanish and $\cE_{\underline{\circ\circ}}$
simplifies to the energy between two spheres \cite{Emig:2007os}. For
an interpretation in terms of multiple scatterings, it is instructive
to use the relation $\ln \det = \tr \ln$ and to expand the logarithm
which yields
\begin{equation}
  \label{eq:energy-2}
  \cE_{\underline{\circ\circ}}  =  - \frac{\hbar c}{2 \pi} \int_0^\infty d\kappa
\sum_{p=1}^\infty \frac{1}{p} \tr \left[
\sum_{n=0}^\infty (-1)^n ( {\mathbb T} {\mathbb U}^I)^n 
  {\mathbb T} ( {\mathbb U}^{21} + {\mathbb U}^{I,21} ) 
 \sum_{n'=0}^\infty (-1)^{n'} ( {\mathbb T} {\mathbb U}^I)^{n'} 
  {\mathbb T} ( {\mathbb U}^{12} + {\mathbb U}^{I,12} )
\right]^p \, ,
\end{equation}
\vspace*{-0.2cm}
\end{widetext}
\vspace*{-1.1cm}
where we have written the inverse matrices of Eq.~\eqref{eq:det-M-1}
as series. The trace acts on an alternating product of T- and
U-matrices which describe scattering and free propagation of EM
fluctuations, respectively. Multiple scatterings between an object and
its image (${\mathbb T} {\mathbb U}^I$) are followed by a propagation
to the other object--image pair, either to the object (${\mathbb
  U}^{21}$) or its image (${\mathbb U}^{I,21}$), between which again
multiple scatterings occur before the fluctuations are scattered back
to the initial object or its image (${\mathbb U}^{12}$ or ${\mathbb
  U}^{I,12}$) and the process repeats. This expansion is useful for small
objects or large separations.


As the first application, we consider the case of two identical,
ground state atoms near a wall, see Fig.~\ref{fig:1}. The separation
between the atoms is $L$ and they have equal distance $H$ from the
wall. In dipole approximation, the retarded limit of the interaction
is described by the static electric ($\alpha_z$, $\alpha_\|$) and
magnetic ($\beta_z$, $\beta_\|$) dipole polarizabilities of the atoms
which can be different perpendicular ($z$) and parallel ($\|$) to the
wall.  The T-matrix of the atoms is diagonal and has finite elements
only for the dipole channel (partial waves with $l=1$), given by
$T^{E}_{10} = \frac{2}{3} \alpha_z \kappa^3$, $ T^{E}_{1m} =
\frac{2}{3} \alpha_\| \kappa^3$ for electric and $T^{M}_{10} =
\frac{2}{3} \beta_z \kappa^3$, $T^{M}_{1m} = \frac{2}{3} \beta_\|
\kappa^3$ for magnetic polarization with $m=\pm 1$. For atoms, the
polarizability is much smaller than $L^3$, and hence it is sufficient
to compute the interaction to second order in the
polarizabilities. This amounts to consider only the terms with $p=1$,
$n=n'=0$ of Eq.~\eqref{eq:energy-2}. The resulting energy
$\cE_{\underline{\circ\circ}}$ is then compared to the well-known
Casimir-Polder (CP) interaction between two atoms (without the wall)
\begin{equation}
  \label{eq:E_CP}
 \cE_{2,|}(L) = -\frac{\hbar c}{8\pi L^7} \!\!
\left[ 33 \alpha_\|^2 +\! 13 \alpha_z^2
- \! 14 \alpha_\|\beta_z + (\alpha \!\leftrightarrow\! \beta) \!\right] 
\end{equation}
which corresponds to the sequence $ {\mathbb T} {\mathbb U}^{21}
{\mathbb T} {\mathbb U}^{12}$ in Eq.~\eqref{eq:energy-2}. The total
interaction energy is
\begin{equation}
  \label{eq:E_CP_plane}
  \cE_{\underline{\circ\circ}}(L,H) = \cE_{2,|}(L) + \cE_{2,\backslash}(D,L) + \cE_3(D,L) 
\end{equation}
with $D=\sqrt{L^2+4H^2}$. The 2-body energy $\cE_{2,\backslash}(D,L)$
comes from the sequence $ {\mathbb T} {\mathbb U}^{I,21} {\mathbb T}
{\mathbb U}^{I,12}$ in Eq.~\eqref{eq:energy-2} and hence is the usual
CP interaction between one atom and the image of the other atom (see
Fig.~\ref{fig:1}).  The change in the relative orientation of the
atoms with $\ell=L/D$ leads to the modified CP potential
\begin{widetext}
\vspace*{-0.7cm}
  \begin{equation}
  \label{eq:E_CP_diag}
  \cE_{2,\backslash}(D,L) = -\frac{\hbar c}{8\pi D^7} \!\!\left[ 26\alpha_\|^2
+\! 20 \alpha_z^2 -\! 14 \ell^2 (4\alpha_\|^2-9\alpha_\|\alpha_z +5\alpha_z^2)
+\! 63\ell^4 (\alpha_\| - \alpha_z)^2  
- 14\!\left(\alpha_\| \beta_\|(1\!-\!\ell^2) +\!\ell^2 \alpha_\| \beta_z \!\right) + (\alpha\!\leftrightarrow \!\beta) \!\right] \, .
\end{equation}
The 3-body energy $\cE_3(D,L)$ corresponds to the sequences
 ${\mathbb T} {\mathbb U}^{21} {\mathbb T}
{\mathbb U}^{I,12}$,  ${\mathbb T} {\mathbb U}^{I,21} {\mathbb T}
{\mathbb U}^{12}$ in Eq.~\eqref{eq:energy-2} and hence describes the
collective interaction between the two atoms and one image atom. 
It is given by
\begin{eqnarray}
  \label{eq:E_3}
  \cE_3(D,L) &=&  \frac{4\hbar c}{\pi} \frac{1}{L^3D^4(\ell+1)^5}\left[ \Big(
3\ell^6 +15\ell^5+28\ell^4+20\ell^3+6\ell^2-5\ell-1\right)\left(\alpha_\|^2-\beta_\|^2
\right)
\nonumber\\ 
&-& \left(3\ell^6+15\ell^5+24\ell^4-10\ell^2-5\ell-1\right) \left(\alpha_z^2-\beta_z^2\right)
+4\left(\ell^4+5\ell^3+\ell^2\right)\left(\alpha_z\beta_\|-\alpha_\|\beta_z
\right)\Big] \, .
\end{eqnarray}
\vspace*{-0.3cm}
\end{widetext}
\vspace*{-1.3cm}
For isotropic electric polarizable atoms this result agrees with that
of Ref.~\cite{Power:1982cs}. It is instructive to consider the two
limits $H\ll L$ and $H\gg L$.  For $H\ll L$ one has $D\to L$ and the
2-body potentials are identical, $\cE_{2,\backslash}(L,L)
=\cE_{2,|}(L)$. The 3-body energy becomes
\begin{equation}
  \label{eq:E_3_small_H}
   \cE_3(L,L) = -\frac{\hbar c}{4\pi L^7} \!\!\left[ -33\alpha_\|^2
+13\alpha_z^2   + 14 \alpha_\| \beta_z - (\alpha\leftrightarrow \beta)\right] \, .
\end{equation}
The total energy $\cE_{\underline{\circ\circ}}$ is now twice the
energy of Eq.~\eqref{eq:E_CP} plus the energy of
Eq.~\eqref{eq:E_3_small_H} and hence $\cE_{\underline{\circ\circ}}$
becomes the CP potential of Eq.~\eqref{eq:E_CP} with the replacements
$\alpha_z\to 2\alpha_z$, $\alpha_\|\to 0$, $\beta_z\to 0$,
$\beta_\|\to 2\beta_\|$, i.e., the 2-body and 3-body contributions add
constructively or destructively, depending on the relative orientation
of a dipole and its image which together form a dipole of zero or twice
the original strength (see Fig.~\ref{fig:fig2}). For $H \gg L$ 
the leading correction to the CP potential of Eq.~\eqref{eq:E_CP} comes
from the 3-body energy which in this limit becomes to order $H^{-6}$
\begin{equation}
  \label{eq:E_3_large_H}
   \cE_3(H,L) = \frac{\hbar c}{\pi} \!\!\left[ \!\frac{\alpha_z^2-\alpha_\|^2}{4 L^3H^4}  +
\frac{9\alpha_\|^2-\alpha_z^2 -2\alpha_\| \beta_z}{8LH^6} - (\alpha\leftrightarrow \beta)\!\right] .
\end{equation} 
The signs of the polarizabilities in the leading term $\sim H^{-4}$
can be understood from the relative orientation of the dipole of one
atom and the image dipole of the other atom, see Fig.~\ref{fig:fig2}.
If these two electric (magnetic) dipoles are almost perpendicular to
their distance vector they contribute attractively (repulsively) to
the potential between the two original atoms. If these electric
(magnetic) dipoles are almost parallel to their distance vector they
yield a repulsive (attractive) contribution. For isotropic
polarizabilities the leading term of Eq.~\eqref{eq:E_3_large_H}
vanishes and the electric (magnetic) part $\sim H^{-6}$ of the 3-body
energy is always repulsive (attractive).

The above results determine the variation of the force between the two
particles with $H$. If the two particles have only either $\alpha_z$
or $\beta_\|$ polarizability, their attractive force is {\it reduced}
when they approach the wall from large $H$ due to the repulsive 3-body
energy. At close proximity to the wall the fluctuations of the dipole
and its image add up to yield a force between the particles that is
{\it enhanced} by a factor of $4$ compared to the force for
$H\to\infty$. Equivalent arguments show that the force between
particles with either $\alpha_\|$ or $\beta_z$ polarizability is {\it
  enhanced} at large $H$ and {\it reduced} to zero for $H\to 0$. This
proves that the force between particles which both have either of the
four polarizabilities is always non-monotonic. The situation can be
different if more than one polarizability is finite, especially for
isotropic particles. In the latter case all contributions (electric,
magnetic, mixed) are enhanced for $H\to 0$ and only the electric term
is reduced at large $H$ so that only the electric part gives a
non-monotonic force. In general, the monotonicity property depends on
the relative strength and anisotropy of the electric and magnetic
polarizabilities.


As the second application we study two macroscopic perfect metallic
spheres of radius $R$ for the same geometry as in the case of atoms
where the lengths $L$ and $H$ are measured now from the centers of the
spheres, see Fig.~\ref{fig:1}. The T-matrix is diagonal and the
elements $T^M_{lm}=(-1)^l \frac{\pi}{2} I_{l+1/2}(\kappa
R)/K_{l+1/2}(\kappa R)$, $T^E_{lm}=(-1)^l \frac{\pi}{2}
[I_{l+1/2}(\kappa R)+2\kappa R I'_{l+1/2}(\kappa R)]
/[K_{l+1/2}(\kappa R)+2\kappa R K'_{l+1/2}(\kappa R)]$ are given in
terms of the Bessel functions $I_\nu$, $K_\nu$. First, we expand the
energy in powers of $R$ by using Eq.~\eqref{eq:energy-2} which implies
that we expand the T-matrices for small frequencies but use the exact
expressions for the U-matrices. For $R \ll L,\, H$ and arbitrary $H/L$
the result for the force can be written as
\vspace*{-.3cm}
\begin{equation}
\vspace*{-.3cm}
  \label{eq:force-of-L}
  F  = \frac{\hbar c}{\pi R^2} \sum_{j=6}^\infty  f_j(H/L) \left(\frac{R}{L}\right)^{j+2} \, .
\end{equation}
The $f_j$ can be computed exactly. We have obtained them up to $j=11$
and the first three are (with $s\equiv \sqrt{1+4h^2}$)
\begin{align}
  \label{eq:h-fcts}
& f_6(h) =  -\frac{1}{16h^8}\Big[s^{-9}(18 + 312 h^2 + 2052 h^4 + 6048 h^6 
\nonumber\\
&\! +  5719 h^8) + 18 - 12 h^2 + 1001 h^8\Big] \, , \quad f_7(h)=0\, , \\
& f_8(h) =  -\frac{1}{160 h^{12}}\Big[s^{-11} (6210 + 140554 h^2 + 1315364 h^4 
\nonumber\\
&\! + 6500242 h^6 + \! 17830560 h^8 + \! 25611168 h^{10} + \! 15000675 h^{12}) \nonumber\\
&\! - 6210 - 3934 h^2 + 764 h^4 - 78 h^6 + 71523 h^{12}\Big] \, .
\end{align}
The coefficient $f_7$ of $R^7$ vanishes since a multipole of order $l$
contributes to the T-matrix at order $R^{2l+1}$ so that beyond the
two-dipole term $\sim R^6$ the next term comes from a dipole ($l=1$)
and a quadrupole ($l=2$), yielding $f_8$. For $H \gg L$ one has
$f_6(h) = -1001/16 +3/(4h^6)+ {\cal O}(h^{-8})$,
$f_8(h)=-71523/160+39/(80h^6)+ {\cal O}(h^{-8})$ so that the wall
induces weak repulsive corrections. For $H \ll L$,
$f_6(h)=-791/8+6741 h^2/8 +{\cal O}(h^4)$, $f_8(h)=-60939/80 + 582879
h^2/80 +{\cal O}(h^4)$ so that the force amplitude decreases  when the spheres are moved a small
distance  away from the wall. This proves the existence of a minimum in
the force amplitude as function of $H/R$ for fixed, sufficiently
small $R/L$. We note that all $f_j(h)$ are finite for
$h\to \infty$ but some diverge for $h\to 0$, e.g., $f_9 \sim f_{11}
\sim h^{-3}$, making them important for small $H$.

\begin{figure}[htbp]
\includegraphics[width=\linewidth]{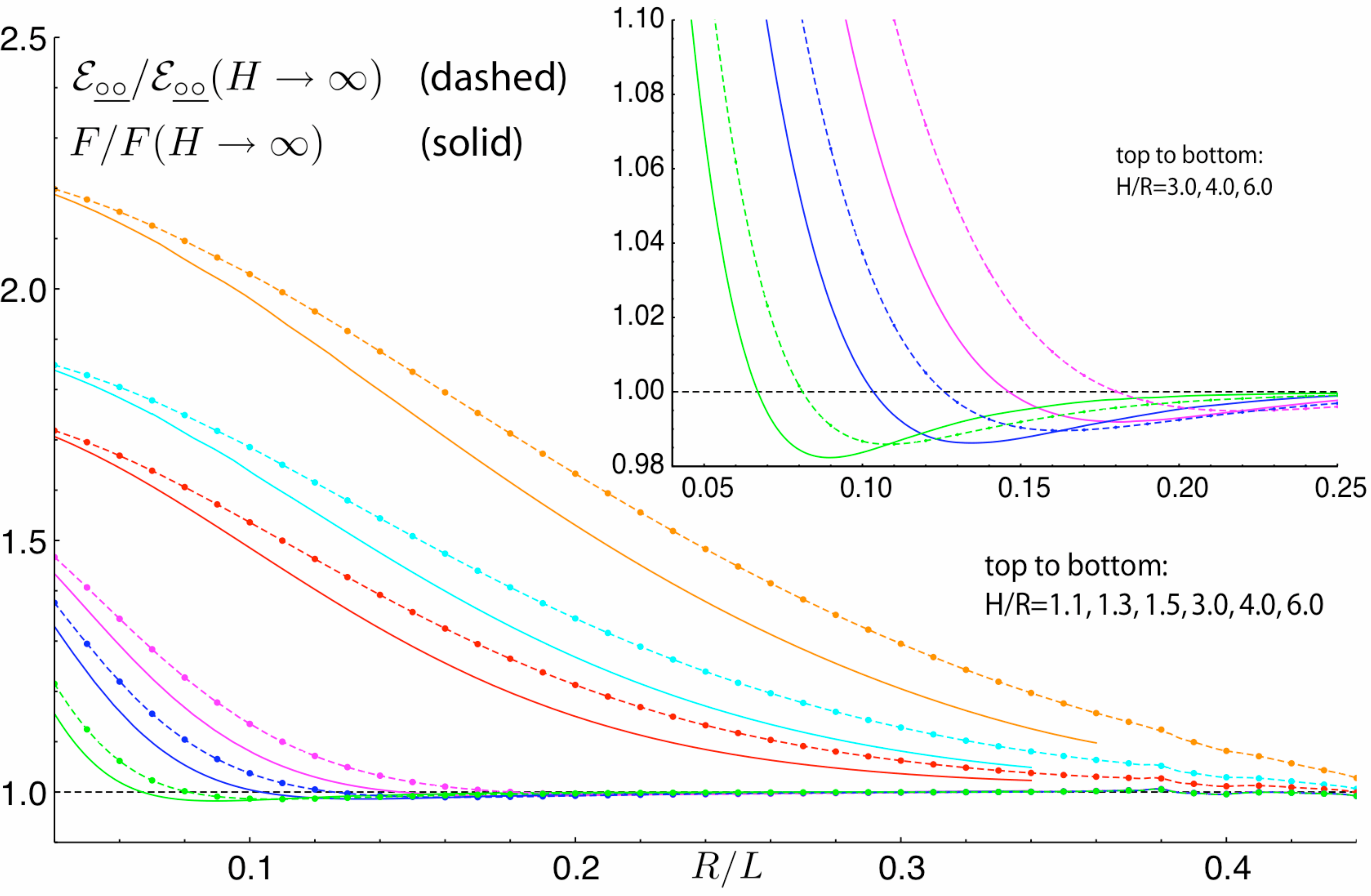}
\vspace*{-.7cm}
\caption{\label{fig:3}Numerical results for the potential energy
  (dashed curves) and force (solid curves) between two spheres as
  function of $R/L$ for different sidewall separations $H/R$. Both
  force and energy are normalized to their values in the absence of
  the sidewall. Inset: Magnification of behavior for small $R/L$.}
\end{figure}
\begin{figure}[htbp]
\vspace*{-.5cm}
\includegraphics[width=\linewidth]{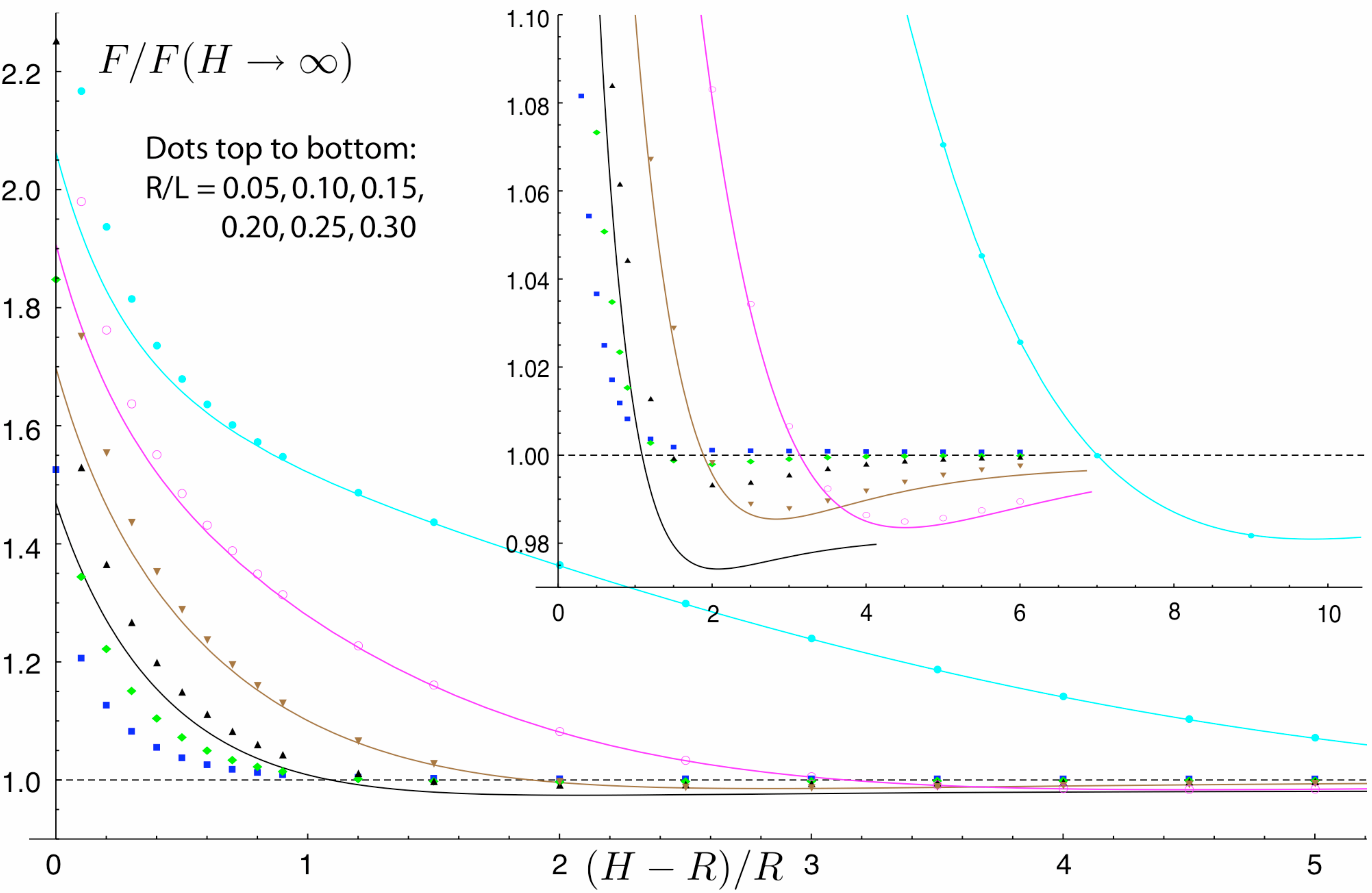}
\vspace*{-.7cm}
\caption{\label{fig:4}Numerical results for the force (dots) between
  two spheres as function of the sidewall separation $H/R$ for
  different sphere separations $R/L$. Shown are also the analytical
  results of Eq.~\eqref{eq:force-of-L}, including terms up to $j=10$
  for $R/L\le 0.2$ (solid curves). Inset: Magnification of the non-monotonicity.}
\vspace*{-.5cm}
\end{figure}

To obtain the interaction at smaller separations or larger radius, we
have computed the energy $\cE_{\underline{\circ\circ}}$ and force
$F=-\partial \cE_{\underline{\circ\circ}} /\partial L$ between the
spheres numerically.  For the energy, we have computed the last
determinant of Eq.~\eqref{eq:det-M-1} and the integral over $\kappa$
of Eq.~\eqref{eq:energy} numerically. The force is obtained by
polynomial interpolation of the data for the energy. The matrices are
truncated at a sufficiently large number of partial waves (with a
maximum truncation order $l_{max}=17$ for the smallest separation) so
that the relative accuracy of the values for
$\cE_{\underline{\circ\circ}}$ is $\approx 10^{-3}$. The data for
$H/R=1$ are obtained by extrapolation in $l_{max}$. The results are
shown in Figs.~\ref{fig:3}, \ref{fig:4}.  In order to show the effect
of the wall, the figures display the energy and force normalized to
the results for two spheres without a wall. Fig.~\ref{fig:3} shows the
energy and force as function of the (inverse) separation between the
spheres for different fixed wall distances. Energy and force show an
increasing relative enhancement due to the wall with increasing $L$,
with the maximal enhancement for small $H$. For sufficiently large
$H/R$, the energy and force ratios are non-monotonic in $L$ and can be
slightly smaller than without the wall. Fig.~\ref{fig:4} shows the
force as function of the wall distance for fixed $L$. When the spheres
approach the wall, the force first decreases slightly if $R/L \lesssim
0.3$ and then increases strongly under a further reduction of $H$. For
$R/L \gtrsim 0.3$ the force monotonically increases towards the
wall. This agrees with the prediction of the large distance
expansion. The expansion of Eq.~\eqref{eq:force-of-L} with $j=10$
terms is also shown in Fig.~\ref{fig:4} for $R/L\le 0.2$. Its validity
is limited to large $L/R$ and not too small $H/R$; it fails completely
for $R/L>0.2$ and hence is not shown in this range.


Our results for atoms are potentially relevant to the interaction
between trapped Bose-Einstein condensates and a surface
\cite{Harber:2005a} at close surface separations. The results for
macroscopic spheres could be important for the design of
nano-mechanical devices where small components operate in close
vicinity to metallic boundaries.  Generally, the wall-induced
enhancement of the interaction can make the experimental observation
of Casimir forces between small particles more feasible. The reported
dependence on the anisotropy of polarizabilities suggest interesting
effects for  objects of non-spherical shape.

We acknowledge helpful discussions with N.~Graham, R.~L.~Jaffe,
and M.~Kardar, and the hospitality of the Institute of
Theoretical Physics, University of Cologne.  This research was
supported by projects MOSAICO, UCM/PR34/07-15859 and a PFU MEC
grant (PR) and by DFG through grant EM70/3 (TE).


\vspace*{-.5cm}

\bibliographystyle{apsrev}
\bibliography{article}

\begin{thebibliography}{9}
\expandafter\ifx\csname natexlab\endcsname\relax\def\natexlab#1{#1}\fi
\expandafter\ifx\csname bibnamefont\endcsname\relax
  \def\bibnamefont#1{#1}\fi
\expandafter\ifx\csname bibfnamefont\endcsname\relax
  \def\bibfnamefont#1{#1}\fi
\expandafter\ifx\csname citenamefont\endcsname\relax
  \def\citenamefont#1{#1}\fi
\expandafter\ifx\csname url\endcsname\relax
  \def\url#1{\texttt{#1}}\fi
\expandafter\ifx\csname urlprefix\endcsname\relax\def\urlprefix{URL }\fi
\providecommand{\bibinfo}[2]{#2}
\providecommand{\eprint}[2][]{\url{#2}}

\bibitem[{\citenamefont{Parsegian}(2005)}]{Parsegian:2005eu}
\bibinfo{author}{\bibfnamefont{V.~A.} \bibnamefont{Parsegian}},
  \emph{\bibinfo{title}{Van der Waals Forces}} (\bibinfo{publisher}{Cambridge
  University Press}, \bibinfo{address}{Cambridge, England},
  \bibinfo{year}{2005}).

\bibitem[{\citenamefont{Power and Thirunamachandran}(1982)}]{Power:1982cs}
\bibinfo{author}{\bibfnamefont{E.~A.} \bibnamefont{Power}} \bibnamefont{and}
  \bibinfo{author}{\bibfnamefont{T.}~\bibnamefont{Thirunamachandran}},
  \bibinfo{journal}{Phys. Rev. A} \textbf{\bibinfo{volume}{25}},
  \bibinfo{pages}{2473} (\bibinfo{year}{1982}).

\bibitem[{\citenamefont{Feinberg}(1974)}]{Feinberg:1974xw}
\bibinfo{author}{\bibfnamefont{G.}~\bibnamefont{Feinberg}},
  \bibinfo{journal}{Phys. Rev. B} \textbf{\bibinfo{volume}{9}},
  \bibinfo{pages}{2490} (\bibinfo{year}{1974}).

\bibitem[{\citenamefont{Emig et~al.}(2007)\citenamefont{Emig, Graham, Jaffe,
  and Kardar}}]{Emig:2007os}
\bibinfo{author}{\bibfnamefont{T.}~\bibnamefont{Emig}},
  \bibinfo{author}{\bibfnamefont{N.}~\bibnamefont{Graham}},
  \bibinfo{author}{\bibfnamefont{R.~L.} \bibnamefont{Jaffe}}, \bibnamefont{and}
  \bibinfo{author}{\bibfnamefont{M.}~\bibnamefont{Kardar}},
  \bibinfo{journal}{Phys. Rev. Lett.} \textbf{\bibinfo{volume}{99}},
  \bibinfo{eid}{170403} (\bibinfo{year}{2007}).

\bibitem[{\citenamefont{Rodriguez et~al.}(2007)\citenamefont{Rodriguez,
  Ibanescu, Iannuzzi, Capasso, Joannopoulos, and Johnson}}]{Rodriguez:2007a}
\bibinfo{author}{\bibfnamefont{A.~W.} \bibnamefont{Rodriguez}},
  \bibinfo{author}{\bibfnamefont{M.}~\bibnamefont{Ibanescu}},
  \bibinfo{author}{\bibfnamefont{D.}~\bibnamefont{Iannuzzi}},
  \bibinfo{author}{\bibfnamefont{F.}~\bibnamefont{Capasso}},
  \bibinfo{author}{\bibfnamefont{J.~D.} \bibnamefont{Joannopoulos}},
  \bibnamefont{and} \bibinfo{author}{\bibfnamefont{S.~G.}
  \bibnamefont{Johnson}}, \bibinfo{journal}{Phys. Rev. Lett.}
  \textbf{\bibinfo{volume}{99}}, \bibinfo{eid}{080401} (\bibinfo{year}{2007}).

\bibitem[{\citenamefont{Rahi et~al.}(2008{\natexlab{a}})\citenamefont{Rahi,
  Rodriguez, Emig, Jaffe, Johnson, and Kardar}}]{Rahi:2008bv}
\bibinfo{author}{\bibfnamefont{S.~J.} \bibnamefont{Rahi}},
  \bibinfo{author}{\bibfnamefont{A.~W.} \bibnamefont{Rodriguez}},
  \bibinfo{author}{\bibfnamefont{T.}~\bibnamefont{Emig}},
  \bibinfo{author}{\bibfnamefont{R.~L.} \bibnamefont{Jaffe}},
  \bibinfo{author}{\bibfnamefont{S.~G.} \bibnamefont{Johnson}},
  \bibnamefont{and} \bibinfo{author}{\bibfnamefont{M.}~\bibnamefont{Kardar}},
  \bibinfo{journal}{Phys. Rev. A} \textbf{\bibinfo{volume}{77}},
  \bibinfo{eid}{030101} (\bibinfo{year}{2008}{\natexlab{a}}).

\bibitem[{\citenamefont{Rahi et~al.}(2008{\natexlab{b}})\citenamefont{Rahi,
  Emig, Jaffe, and Kardar}}]{Rahi:2008kb}
\bibinfo{author}{\bibfnamefont{S.~J.} \bibnamefont{Rahi}},
  \bibinfo{author}{\bibfnamefont{T.}~\bibnamefont{Emig}},
  \bibinfo{author}{\bibfnamefont{R.~L.} \bibnamefont{Jaffe}}, \bibnamefont{and}
  \bibinfo{author}{\bibfnamefont{M.}~\bibnamefont{Kardar}},
  \bibinfo{journal}{Phys. Rev. A} \textbf{\bibinfo{volume}{78}},
  \bibinfo{eid}{012104} (\bibinfo{year}{2008}{\natexlab{b}}).

\bibitem[{\citenamefont{Emig}(2008)}]{Emig:2008ee}
\bibinfo{author}{\bibfnamefont{T.}~\bibnamefont{Emig}},
  \bibinfo{journal}{Journal of Statistical Mechanics: Theory and Experiment}
  \textbf{\bibinfo{volume}{4}}, \bibinfo{pages}{P04007} (\bibinfo{year}{2008}).

\bibitem[{\citenamefont{Harber et~al.}(2005)\citenamefont{Harber, Obrecht,
  McGuirk, and Cornell}}]{Harber:2005a}
\bibinfo{author}{\bibfnamefont{D.~M.} \bibnamefont{Harber}},
  \bibinfo{author}{\bibfnamefont{J.~M.} \bibnamefont{Obrecht}},
  \bibinfo{author}{\bibfnamefont{J.~M.} \bibnamefont{McGuirk}},
  \bibnamefont{and} \bibinfo{author}{\bibfnamefont{E.~A.}
  \bibnamefont{Cornell}}, \bibinfo{journal}{Phys. Rev. A}
  \textbf{\bibinfo{volume}{72}}, \bibinfo{pages}{033610}
  (\bibinfo{year}{2005}).

\end{thebibliography}

\end{document}